\pdfoutput=1
\documentclass[twocolumn,nofootinbib]{revtex4}
\usepackage{amsmath}
\usepackage{amsfonts}
\usepackage{amssymb}
\usepackage{enumerate}
\usepackage{braket}
\usepackage{xcolor}
\usepackage{hyperref}

\newcommand\paper{Letter{}}
\newcommand\myendnote\footnote

\DeclareMathOperator{\tr}{tr}

\renewcommand{\d}{{\rm d}}

\newcommand\ithree{{I_3}}

\newcommand\nhat{{{\hat{\bf n}}}}

\newcommand\nhatone{{{\hat{\bf n}_1}}}
\newcommand\nhattwo{{{\hat{\bf n}_2}}}
\newcommand\nhatthree{{{\hat{\bf n}_3}}}
\newcommand\nhatfour{{{\hat{\bf n}_4}}}

\newcommand\rhoww{{\rho}}
\newcommand\rhow{{\rho_{W}}}

\newcommand\chshlabel{{\rm gen}}

\newcommand\bcglmpxy{{\mathcal{B}_{\rm CGLMP}^{\prime\,xy} }}
\newcommand\itwomax{{\mathcal{I}_2^{\chshlabel,\max}}}
\newcommand\bcglmpswapxy{{\mathcal{B}_{\rm CGLMP}^{xy} }}
\newcommand\bcglmpswapyz{{\mathcal{B}_{\rm CGLMP}^{yz} }}
\newcommand\bcglmpswapzx{{\mathcal{B}_{\rm CGLMP}^{zx} }}

\newcommand\ithreemax{{\mathcal{I}_3^{\rm xyz}}}

\newcommand\intlumi{{\textrm{140\,fb}^{-1}}}

\begin{document}


\title{Testing Bell inequalities in Higgs boson decays}
\date{First submitted: 2 June 2021. This version: \today}
\author{Alan J. Barr}
\affiliation{Department of Physics, Keble Road, University of Oxford, OX1 3RH}
\affiliation{Merton College, Merton Street, Oxford, OX1 4JD}


\begin{abstract}
Higgs boson decays produce pairs of $W$ bosons in a maximally entangled state,
the spins of which can be expected to violate Bell inequalities.
We show that the spin density matrix of the $W^\pm$ pair may be reconstructed experimentally
from the directions of the charged lepton decay products, and from it the expectation values of various Bell operators determined. 
Numerical simulations of $H\rightarrow WW^*$ decays indicate that violation of a generalised CHSH inequality is unlikely to be measurable, 
however the CGLMP inequality is near-maximally violated.
Experimental Bell tests could be performed at a variety of colliders and in different production channels.
If reconstruction effects and backgrounds can be controlled then statistically significant violations could be observable even with datasets comparable to those already collected at the LHC.
\end{abstract}
\maketitle


\section{Introduction}

The predictions of quantum mechanics for entangled particles 
have long been known to violate inequalities of the type first introduced by Bell~\cite{BellOnTheEPR}. 
Violations of such inequalities are expected in quantum theory, 
however they are incompatible with `realist' theories (including classical physics) 
in which the properties of systems are independent of our observation of them.

Experimental tests showing violation of Bell inequalities have been performed
for pairs of two-outcome measurements using
photons~\cite{PhysRevLett.28.938,PhysRevLett.49.1804}, ions~\cite{Rowe}, superconducting systems~\cite{josephson} and nitrogen vacancy centres~\cite{pfaff}, 
and in pairs of three-outcome measurements using photons~\cite{PhysRevLett.89.240401}.

Significant advances in testing local realism were made through ``loophole-free'' Bell tests performed by three different groups in 2015~\cite{Hensen:2015ccp,Giustina_2015,Shalm_2015}. 
Proposals have also been made to test Bell inequalities in $e^+e^-$ collisions~\cite{Tornqvist}, charmonium decays~\cite{Baranov_2008,10.1093/ptep/ptt032} and positronium decays~\cite{Ac_n_2001}.
Recently it has been proposed to make such tests in entangled $t+\bar{t}$ decays~\cite{Afik_2021,Fabbrichesi:2021npl}
and in systems of $B^0 + \bar B^0$ mesons~\cite{2106.07399} at the LHC.\footnote{Following the submission of this \paper{} a manuscript was submitted to the arXiv exploring sensitivity to entanglement and to Bell inequalities in $t+\bar{t}$ decays at the LHC~\cite{Severi:2021cnj}.}
\par
The decay $H\rightarrow WW^*$ is that of a scalar 
to a pair of distinguishable massive spin-one bosons in a maximally entangled state.
In the narrow width and non-relativistic approximations the state may be represented in the spin basis as
\begin{equation}\label{eq:singlet}
\ket{\psi_{s}} = \tfrac{1}{\sqrt{3}} \big( \ket{+}\ket{-} - \ket{0} \ket{0} + \ket{-}\ket{+} \big).
\end{equation}

The subsequent $W$ boson decays maximally violate chirality;
a $W^+$ boson decay preferentially emits a charged lepton along the $W^+$ spin direction
while a $W^-$ boson decay preferentially emits a charged lepton moving against its spin direction. 
Decaying $W$ bosons are ``their own polarimeters''~\cite{Tornqvist},
with each decay causing a spin measurement to be made along the axis of the emitted lepton. 
This results in correlations in the azimuthal directions of the emitted
leptons $\ell^\pm\in\{e^\pm,\mu^\pm\}$ 
which were exploited by the ATLAS and CMS collaborations to separate Higgs decays 
from $W^+W^-$ backgrounds in their Higgs boson searches~\cite{Aad_2012,Chatrchyan_2012}.
\par
Moreover, measurements of the emitted lepton directions over an ensemble of 
decays allow one to determine the two-particle spin density matrix $\rhoww$,
and from it the expectation values $\tr(\rhoww\mathcal{B})$
of various quantum Bell operators $\mathcal{B}$~\cite{PhysRevLett.68.3259}. 
One can therefore go further and use these $H\rightarrow WW^*$ boson decays 
as a laboratory to perform tests of Bell inequalities.
\par
Like similar proposals using the weak decay to analyse spin~\cite{Tornqvist,Baranov_2008,Afik_2021,Fabbrichesi:2021npl}
they do not allow the experimentalist to freely choose the $W$-boson spin measurement directions.
Nevertheless, they offer an opportunity to test Bell inequalities in a new regime:
at energies of order $m_H \approx 125\,\textrm{GeV}$~\cite{ParticleDataGroup:2020ssz},
time scales of order $\hbar/\Gamma_W \approx 10^{-25}\,\rm{s}$~\cite{ParticleDataGroup:2020ssz}
and length scales of order $\hbar c/\Gamma_W \approx 10^{-16}\,\rm{m}$. 
These scales are very many orders of magnitude removed from existing experimental results, 
and offer the prospect of performing Bell tests in a new regime deep within the realm of quantum field theory.


\section{Testing the generalised CHSH inequality}\label{sec:chsh}

For a pair of two-outcome experiments, such as measurements of the spins of a pair of spin-half particles,
the necessary and sufficient conditions~\cite{fine_prl} that measurements be compatible with any realist theory is that 
they satisfy the inequality of Clauser-Horne-Shimony-Holt (CHSH)~\cite{chsh} 
\begin{equation}\label{eq:chsh}
\mathcal{I}_2 =E(a, b)-E(a, b') + E(a', b) + E(a', b') \leq 2\,.
\end{equation}
This equation characterises the results $a$, $b$ of experiments performed on each of the systems, labelled A and B respectively, 
where the primes indicate the results obtained from alternative versions of the experiment, for example
by changing the axis of a spin measurement. 
\par
For two-outcome measurements such as of the spins of spin-half particles or of photons, 
the expectation values $E$ are the products of the assigned values, either $+1$ or $-1$. 
If one demands a description of nature which is consistent with realism
then the values of $\mathcal{I}_2$ can be no larger than two,
since a larger value would imply negative marginal probabilities.
\par
Theories admitting local realism are based on the further observation that according to special relativity information should not travel faster than the speed of light. In local realistic theories the hypothesised pre-existing values for the experimental outcomes for each system may therefore depend only on events in the past light-cone of that system. This restriction prevents information from the measurement settings or measurements of system A from being transmitted to B, and hence to causally affect the measurement outcome at B (and similarly for signals from B to A). 
\par
Quantum mechanics, in conflict both with realism and with local realism, 
allows values of $\mathcal{I}_2$ larger than two and indeed 
up to the Cirel'son bound~\cite{cirelson} of $2\sqrt{2}$.
\par
For the case of massive vector bosons
we have to assign a result value to each of the three possible outcomes from the measurements,
opening up more assignment options. 
Here we choose to take the eigenvalues $\{1,0,-1\}$ of the $s=1$ spin operators meaning that the additional
outcome is assigned the value zero~\cite{Caban_2008}. 
This additional zero outcome dilutes the expectation values and so tends to 
decrease violation of inequality~\eqref{eq:chsh}, such that 
this spin-one generalisation of the spin-half CHSH operator is found in analytical calculations 
not to violate the Bell inequality \eqref{eq:chsh}
either in non-relativistic quantum mechanics or in relativistic quantum mechanics in the narrow-width approximation \cite{Caban_2008}. Nevertheless, it proves instructive to examine its behaviour in the $H\rightarrow WW^*$ system 
to allow insight into its construction and behaviour and so that a direct comparison with the spin-half equivalent measurement may be made. 

The generalised CHSH operator that we use~\cite{Caban_2008} may be written  
\begin{equation}\label{eq:chshoperatorspin}
\mathcal{B}_{\rm CHSH}^{\chshlabel} =\nhatone \cdot \mathbf{S} \otimes (\nhattwo - \nhatfour)\cdot {\bf S}
+ \nhatthree \cdot \mathbf{S} \otimes (\nhattwo + \nhatfour )\cdot  {\bf S}\,,
\end{equation}
in an analogous manner to the spin-half case, 
where $\nhatone$, $\nhattwo$, $\nhatthree$ and $\nhatfour$
are unit vectors in $\mathbb{R}^3$ and $\mathbf{S}\equiv(S_x,S_y,S_z)$ are the dimension-3
Hermitian spin operators. 
When an explicit matrix representation is required we use the standard representation i.e. with $S_z={\rm diag}(+1,0,-1)$.

To calculate the expectation value
\begin{equation}\label{eq:chshmean}
\mathcal{I}_2^{\chshlabel} = \tr(\rhoww {\mathcal{B_{\rm CHSH}^{\chshlabel}}}),
\end{equation}
we note that the density matrix $\rho$ for a single spin-one particle may be parameterised by
\begin{equation}\label{eq:rhoone}
\rhow =\tfrac{1}{3} \ithree + \sum_{i=1}^3 a_i S_i + \sum_{i,j=1}^3 c_{ij} S_{\{ij\}}, 
\end{equation}
where we denote the anticommutator 
\begin{equation*}
S_{\{ij\}} \equiv S_i S_j + S_j S_i,
\end{equation*}
and where the parameters $a_i$ form a real vector and $c_{ij}$ a traceless real symmetric matrix.

The two-particle spin density matrix $\rhoww$ may similarly be parameterised in terms of the $S_i$ and $S_{\{ij\}}$ for each particle.  
Noting that the $S_i$ are each trace orthogonal with one another, with the identity, and with each of the $S_{\{ij\}}$,
and that the spin operators of each particle commute with those of the other, the only terms in $\rhoww$ contributing to 
the expectation value~\eqref{eq:chshmean} are of the form
\begin{equation}\label{eq:rhocontainsdij}
\rhoww \supset \sum_{i,j=1}^3 \tfrac{1}{4}d_{ij}  \,S_i \,\otimes  S_j ,
\end{equation}
where $d_{ij}$ are real parameters which contribute to this expectation value through terms of the form
\begin{equation}\label{eq:trsisj}
\tr(\rhoww S_i\otimes S_j) = d_{ij}.
\end{equation}
Our generalised CHSH inequality for a pair of spin-one particles therefore can be reduced to 
\begin{equation}\label{eq:chshoperatorvectorized}
\left| \nhatone \cdot d \cdot (\nhattwo-\nhatfour) + \nhatthree \cdot d \cdot (\nhattwo+\nhatfour)\right| \leq2.
\end{equation}
\par
We next need to determine the elements $d_{ij}$ of the matrix $d$ and to choose four unit vectors
that maximise the left hand side of \eqref{eq:chshoperatorvectorized}.
In practice not all choices need be made, since a procedure for testing the inequality in general 
has been obtained for the spin-half case~\cite{HORODECKI1995340}, 
and remains valid for a pair of spin-one particles. 

Starting from the real matrix $d$ and its transpose $d^T$ one forms the real symmetric positive matrix $M\equiv d^Td$.
One orders the three eigenvalues  $\mu_1,\,\mu_2,\,\mu_3$ of $M$ such that $\mu_1\geq \mu_2 \geq \mu_3$.
The largest value of our generalised CHSH operator is then, following~\cite{HORODECKI1995340},
\begin{align}\label{eq:chshmax}
\itwomax &\equiv \max_{\{\nhatone,\nhattwo,\nhatthree,\nhatfour\}} \left( \braket{\mathcal{B}_{\rm CHSH}^{\chshlabel}} \right)  \nonumber\\
               & = 2 \sqrt{\mu_1 + \mu_2}.
\end{align}
This value may be compared against the bound~\eqref{eq:chsh} required of a realist theory.

We note that while we choose to make a spin-eigenvalue-based assignment for the outcomes, there are other  operators which make different assignments of states to  outcomes~\cite{Mermin:PhysRevD.22.356,PhysRevLett.49.901,Mermin:PhysRevLett.65.1838,GISIN199215,PhysRevA.65.032118,Collins_2002}, and for which the quantum mechanical expectation values, unlike \eqref{eq:chshoperatorspin},  do violate the inequalities implied by realist theories. 
We will not investigate all of these possibilities in this \paper{}, but rather in Section~\ref{sec:qutrit} we investigate just the tightest of them. 


\section{Testing the CGLMP inequality}\label{sec:qutrit}

The optimal~\cite{masanes} Bell inequality for pairs of three-outcome systems is 
the Collins-Gisin-Linden-Massar-Popescu (CGLMP) inequality~\cite{PhysRevA.65.032118,Collins_2002}. 
To construct it one again considers two observers $A$ and $B$, 
each having two measurement settings,  $A_1$ and $A_2$ for $A$, and $B_1$ and $B_2$ for $B$,
but with each experiment now having three possible outcomes.
One denotes by $P(A_i=B_j+k)$ the probability that the outcomes $A_i$ and $B_j$
differ by $k$ modulo $3$. One then constructs the linear function
 \begin{multline}\label{eq:cglmp}
 \mathcal{I}_3 = P(A_1=B_1) + P(B_1=A_2+1) + P(A_2=B_2) \\ 
 + P(B_2=A_1) - P(A_1=B_1-1) - P(B_1=A_2)  \\- P(A_2=B_2-1) - P(B_2=A_1-1).
 \end{multline}
In classical theories, and other theories admitting realism, this function is bounded by~\cite{Collins_2002}
\begin{equation}\label{eq:cglmpbound}
\mathcal{I}_3 \leq 2.
\end{equation}

To test inequality \eqref{eq:cglmpbound} in quantum mechanics we can calculate the expectation value of the Bell operator 
\begin{multline}\label{eq:cglmpmixed}
\bcglmpswapxy =  - \tfrac{2}{\sqrt{3}} \left( S_x \otimes S_x + S_y \otimes S_y \right) \\
                                               + \lambda_4\otimes\lambda_4 + \lambda_5\otimes\lambda_5 ,
\end{multline}
where $\lambda_i$ is the $i$th of the eight traceless $3\times3$ Hermitian Gell-Mann matrices
in the standard representation~\cite{PhysRev.125.1067}. 
In this convention the spin operators $S_x$ and $S_y$
are given by
\[
S_x = \tfrac{1}{\sqrt{2}}(\lambda_1 + \lambda_6) \quad \textrm{and} \quad S_y = \tfrac{1}{\sqrt{2}}(\lambda_2 + \lambda_7).
\]

Our operator \eqref{eq:cglmpmixed} is related to the standard CGLMP operator $\bcglmpxy$~\cite{Acin_2002}
through the transformation\myendnote{A similar permutation operation is also required in principle for the
generalised CHSH operator, but has no net effect on the final result after the process of optimisation over measurement directions.}
\begin{equation}
\bcglmpswapxy= ( T \otimes \ithree) \bcglmpxy( T \otimes \ithree),
\end{equation}
where the operator $T$ has non-zero elements $(1,-1,1)$ on the secondary diagonal.
This procedure has the same effect as mapping our singlet state \eqref{eq:singlet} into the computational basis:
\[
\ket{\psi_s}\rightarrow \tfrac{1}{\sqrt{3}}(\ket{0}\ket{0}+\ket{1}\ket{1}+\ket{2}\ket{2}).
\]


\section{Determining expectation values from data}\label{sec:fromdata}

We wish to determine  the expectation value $\tr(\rhoww\mathcal{B})$ of two different Bell operators $\mathcal{B}$ from  $H\rightarrow WW^*$ decay data. 
We may do so by finding the density matrix $\rho$ using as data the directions $\nhat_{\ell^+}$ and $ \nhat_{\ell^-}$ of the daughter leptons. 
Exploiting the trace orthogonality relations
\begin{equation}\label{eq:GMtrace}
\tr(\lambda_i \lambda_j) = 2\delta_{ij},
\end{equation}
we now parameterise the density matrix for the $W^+W^-$ spins in the Gell-Mann basis
\begin{multline}\label{eq:rhoGM}
\rhoww = \tfrac{1}{9} \ithree \otimes \ithree + \sum_{i=1}^8 f_i \lambda_i \otimes \ithree + \sum_{j=1}^8 g_j \ithree \otimes \lambda_j \\
 + \sum_{i,j=1}^8 h_{ij} \lambda_i \otimes \lambda_j ,
\end{multline}
where $f_i$, $g_i$ and $h_{ij}$ are real coefficients, of which only the $h_{ij}$ contribute to the Bell operators.

As a preliminary, let us consider the spin density matrix 
\begin{equation}\label{eq:rhooneGM}
\rhow = \tfrac{1}{3} \ithree +\sum_{i=1}^8 \Lambda_i \lambda_i ,
\end{equation}
for a single $W^+$ or $W^-$ boson, where the $\Lambda_i$ are real coefficients.
The probability density function for a $W^\pm$ boson with 
spin density matrix given by \eqref{eq:rhooneGM} to emit a charged lepton $\ell^\pm$ 
into infinitesimal solid angle $\d\Omega$ in the direction $\hat{\bf n}(\theta,\phi)$ is 
\begin{equation}\label{eq:definepdf}
p(\ell^\pm_{\hat{\bf n}}; \rhow) = \tfrac{3}{4\pi} \tr(\rhow \Pi_{\pm,\hat{\bf n}}) ,
\end{equation}
where  $\Pi_{\pm,\hat{\bf n}}\equiv \ket{\pm}_{\nhat}\bra{\pm}_{\nhat}$ are projection operators\myendnote{The positive helicity $\ell^+$ travelling in direction $\hat{\bf n}$
in the $W^+$ rest frame is accompanied by a negative helicity $\nu_\ell$ travelling in the opposite direction $-\hat{\bf n}$, 
and similarly the negative helicity $\ell^-$ with a positive helicity $\nu_\ell$ in the $W^-$ frame, so these are spin-one projections.}, 
the roles of which are to select negative helicity $\ell^-$ or positive helicity $\ell^+$ 
in the direction $\hat{\bf n}$.
The normalisation of~\eqref{eq:definepdf} is such that  
$\int\d\Omega \, p(\ell^\pm_{\hat{\bf n}}; \rhow) = 1$.
\par
Using \eqref{eq:definepdf} we may obtain information about the density matrix parameters
$\Lambda_i$ from angular integrals. In particular
\begin{align}
\braket{\xi_x^\pm}_{\rm av} & = \int\d\Omega\,  p(\ell^\pm_{\hat{\bf n}}; \rhow) \sin\theta \cos\phi \nonumber\\
                                            & = \pm \tfrac{1}{\sqrt{2}} (\Lambda_1 + \Lambda_6 ) \label{eq:Ax}, \\
\braket{\xi_y^\pm}_{\rm av} & = \int\d\Omega\,  p(\ell^\pm_{\hat{\bf n}}; \rhow) \sin\theta\sin\phi    \nonumber\\
                                            & =  \pm \tfrac{1}{\sqrt{2}} (\Lambda_2 + \Lambda_7 ) \label{eq:Ay}, \\
\braket{\xi_z^\pm}_{\rm av} & = \int\d\Omega\,  p(\ell^\pm_{\hat{\bf n}}; \rhow) \cos\theta    \nonumber\\
                                            & =  \pm \tfrac{1}{2} (\Lambda_3+ \sqrt{3} \Lambda_8 ) \label{eq:Az},
\end{align}
where the direction cosines
$\xi_i^+ = \nhat_i \cdot \nhat_{\ell^+}$ and $\xi^-_j = \nhat_j \cdot \nhat_{\ell^-}$
are measured in the rest frames of the $W^+$ and $W^-$ bosons respectively.
Equations \eqref{eq:Ax}--\eqref{eq:Az} 
allow us to determine the expectation values 
\[
\tr(\rhow S_i) = \pm 2 \braket{\xi_i^\pm}_{\rm av}
\]
of the single-particle spin operators from the data. 
\par
Extending the calculation to the two-particle density matrix we can calculate the expectation 
values of the operators required for the generalised CHSH inequality in terms of observables\myendnote{We note that the factor of 4 in \eqref{eq:sitimessj} differs from a factor of 9 which would be obtained in the spin-half case.
}:
\begin{equation}\label{eq:sitimessj}
\tr(\rhoww S_i \otimes S_j) = -4 \braket{\xi^-_i \xi^+_j}_{\rm av}.
\end{equation}
In the absence of experimental cuts, and provided the sample of events is sufficiently large\myendnote{Care 
is necessary in evaluating $\itwomax$ when event samples become very small, 
due to the procedure of maximising over the choice of axes. 
In the limit where only a single event satisfies the selection requirements 
there exists a choice of axes for which $\xi_{i}^+=\xi_j^-=1$, 
hence $\mu_1+\mu_2=16$ and so the inferred value of $\itwomax$ would be $2\sqrt{16}=8$.},  
the elements of $d$ in \eqref{eq:rhocontainsdij} are therefore given by 
\begin{equation}
d_{ij} 
 = - 4 \braket{\xi_i^+\xi^-_j}_{\rm av}\label{eq:measurecorrelation},
\end{equation}
from which we may calculate the generalised CHSH 
inequality \eqref{eq:chshoperatorvectorized} for any measurement angles.

\par
The CGLMP expectation value in terms of the parameters of $\rhoww$ is
\begin{multline}\label{cglmpcoefficients}
\tr(\rhoww \bcglmpswapxy) = 
4 (h_{44} +  h_{55}) 
 - \tfrac{4}{\sqrt{3}} ( h_{11} + h_{16} +  h_{61} + h_{66} ) \\
 - \tfrac{4}{\sqrt{3}} ( h_{22} + h_{27} +  h_{72} + h_{77} ) .
\end{multline}
The terms in the second and third parentheses come from $S_x\otimes S_x$ and $S_y\otimes S_y$
operators respectively so  can be determined using \eqref{eq:sitimessj}.
\par
To determine the remaining terms we return to the single-particle density matrix \eqref{eq:rhooneGM}, and note that the angular integrals
\begin{align}
\braket{(\xi_x^\pm)^2-(\xi_y^\pm)^2}_{\rm av} & = \int\d\Omega\,  p(\ell^\pm_{\hat{\bf n}}; \rhow) \sin^2\theta\cos(2\phi) \nonumber\\
                                            & = \tfrac{2}{5} \Lambda_4 \label{eq:A4}
\end{align}
and
\begin{align}
2 \braket{\xi_x^\pm \xi_y^\pm }_{\rm av} & = \int\d\Omega\,  p(\ell^\pm_{\hat{\bf n}}; \rhow) \sin^2\theta\sin(2\phi) \nonumber\\
                                            & = \tfrac{2}{5} \Lambda_5 \label{eq:A5},
\end{align}
extract the parameters of interest, so that the expectation values are
\begin{align}
\tr( \rhow \lambda_4 ) &= 5 \left\langle (\xi_x^\pm)^2 - (\xi^\pm_y)^2 \right\rangle_{\rm av} \nonumber\\
\tr ( \rhow \lambda_5 )  &= 10\left\langle \xi_x^\pm \xi_y ^\pm \right\rangle_{\rm av}. \label{eq:trrholambda}
\end{align}

Extending \eqref{eq:trrholambda} to the two-particle density matrix, 
the CGLMP expectation value can be expressed
\begin{multline}\label{eq:cglmpest}
\tr(\rhoww \bcglmpswapxy ) = 
\tfrac{8}{\sqrt{3}}\left\langle\xi^+_x \xi^-_x + \xi^+_y \xi^-_y\right\rangle_{\rm av} \\
+25 \left\langle \left( (\xi_x^+)^2 - (\xi^+_y)^2\right) \left( (\xi_x^-)^2 - (\xi_y^-)^2 \right) \right\rangle_{\rm av} \\
+ 100\left\langle \xi_x^+\xi_y^+\xi^-_x\xi^-_y \right\rangle_{\rm av}.
\end{multline}
in terms of the $x$- and $y$-direction cosines of the lepton emission directions in the respective $W^\pm$ boson rest frames.
This is our main result, and provides an experimental observable that can be used to test 
the CGLMP Bell operator experimentally against the classical bound 
in any process in which Higgs bosons are produced and subsequently decay to $WW^*$.

The expectation value in \eqref{eq:cglmpest} is calculated using only $x$ and $y$ axis direction cosines.
Corresponding operators and expectation values could also be constructed for other 
pairs of mutually orthogonal axes, 
and tested against the CGLMP inequality. 
Each such rotated operator will have its own particular dependence
on the two-particle density matrix parameters. 
The expectation value of the CGLMP operator 
for the new axes can be calculated from a generalisation 
of Eqn.~\eqref{eq:cglmpest} but now using the corresponding direction cosines.
In the case of $H\rightarrow WW^*$ decays the ensemble of decays has rotational symmetry in the Higgs boson rest frame
around the direction of the W boson momenta. 
Hence in this \paper{} rather than testing every possible pair of axes\myendnote{This procedure does not formally exclude the possibility that there exists another pair of orthogonal axes which would show larger violation in the general case, but it is sufficient for our purposes.}
we choose a set of Cartesian coordinates in which one axis is aligned with this privileged direction, 
construct expectation values for each of the $(x,y)$, $(y,z)$ and $(z,x)$ pairs of axes, and compare the 
largest of them
\begin{equation}\label{eq:cglmpmax}
\ithreemax  = {\rm max} \left ( \braket{\bcglmpswapxy},\braket{\bcglmpswapyz},\braket{\bcglmpswapzx} \right)
\end{equation}
to the classical bound~\eqref{eq:cglmpbound}.


\section{Numerical simulations}\label{sec:numerical}

In the non-relativistic and narrow-width limits, a measurement of the CGLMP operator
for the spin-singlet state of a pair of $W$ bosons from a Higgs boson decay is expected to violate the corresponding Bell inequality~\eqref{eq:cglmpbound},
since the problem reduces to the non-relativistic quantum mechanical calculation~\cite{Collins_2002}.
To investigate the impact of relativistic and finite-width effects a numerical calculation is employed. 
These numerical simulations will also allow us to perform a first investigation of the impact of 
experimental resolutions and event selections, in particular
for the existing general-purpose detectors at the LHC. 

We performed Monte Carlo simulations of $gg \rightarrow H \rightarrow \ell^+ \nu_\ell \,\ell^-{\bar{\nu}}_\ell$
events, where $\ell \in \{e,\,\mu\}$, using the {\texttt Madgraph v2.9.2} software~\cite{Alwall_2014}  which includes
full spin correlation, relativistic and Breit-Wigner effects. 
A sample of $10^6$  events was generated at leading order at a proton-proton centre-of-mass energy of 13\,TeV,
using the Higgs effective-field theory model.
Higher order corrections to the shapes of the normalised angular distributions, 
which for Higgs boson decays to four leptons are typically at the $\lesssim$\,5\% level~\cite{Prophecy4f2006,Boselli:2015aha,Prophecy4f2015},
are neglected in this initial study. 
The LHC Higgs cross-section working group has calculated the $gg\rightarrow H$ cross section for a 125\,GeV Higgs boson 
to N$^3$LO in the effective theory to be 48.6\,pb~\cite{hxswg4}.
The branching ratio $H\rightarrow\ell^+ \nu \ell^{\prime-}  \bar\nu$ for $\ell,\ell^\prime \in \{e,\,\mu\}$ has been also 
calculated to be $1.055\times10^{-2}$~\cite{hxswg4}.
Thus our sample of $10^6$ events corresponds to an integrated luminosity of 1950\,fb$^{-1}$.
Using these values, the simulations were scaled to the target integrated luminosity of $\intlumi$, 
approximately that recorded by each of the ATLAS and CMS experiments during the period 2015--2018~\cite{boyd2020lhc}. 
Events containing an $e^+e^-$ or $\mu^+\mu^-$ pair were rejected 
in order to remove $H\rightarrow ZZ^{*}$ contributions. 
\par
Our choice of orthonormal basis for the matrix $d$ is a modification of that
proposed for measuring spin correlation in top quarks~\cite{Bernreuther_2015}. 
In the $W^+W^-$ centre-of-mass frame the direction 
of the $W^+$ is denoted $\hat{\bf k}$. The direction $\hat{\bf  p}$ of one of the beams in that frame 
is determined, and a mutually orthogonal basis constructed from them:
\[
\hat{\bf k},  \qquad 
\hat{\bf r} = \frac{1}{r}(\hat{\bf p}-y \hat{\bf{k}}), \qquad 
\hat{\bf n} = \frac{1}{r}(\hat{\bf p} \times \hat{\bf k}),
\]
where $y=\hat{\bf p} \cdot \hat{\bf k}$ and $r=\sqrt{1-y^2}$.
This provides a right-handed orthonormal basis $\{\hat{\bf n},\,\hat{\bf r},\,\hat{\bf k}\}$  
defined in the Higgs boson rest frame. Boosts are then performed
into each of the $W^\pm$ rest frames, 
and a new basis $\{\hat{\bf x},\,\hat{\bf y},\,\hat{\bf z}\} = \{\hat{\bf n},\,\hat{\bf r},\,\hat{\bf k}^\prime\}$ 
defined in each such that $\hat{\bf n}$ and $\hat{\bf r}$ are unmodified, while 
each $\hat{\bf k}^\prime$ is parallel to $\hat{\bf k}$ but has been unit-normalised after the corresponding boost. 
The correlation matrix $d$ is then constructed according to~\eqref{eq:measurecorrelation},
and the CGLMP expectation values according to \eqref{eq:cglmpest} and \eqref{eq:cglmpmax}.

Since $m_H<2m_W$, at least one of the $W$ bosons must be off its mass-shell,
and therefore can be expected to have some scalar component. This component will behave like noise, reducing the observed correlations, 
so we might expect the degree of the observed correlations in the simulation to depend on the range of the $W$ boson masses accepted.
Changing the selected range of $W^*$ boson masses can also be expected to modify the impact of relativistic and finite-width corrections. 
The values of $\itwomax$ and $\ithreemax$ were therefore determined for several different selections, 
each being defined by the veto on same-flavour leptons and a lower bound $m_W^<$ on the smaller of the masses of the two reconstructed $W$ bosons.

\begin{table}
\setlength{\tabcolsep}{6pt}
\begin{tabular}{ l  c c  c  c c} 
\hline\hline
$m_W^<$~[GeV]	&	0	& 20  	&	30		& 40		& 50		\\
\hline\hline
$\itwomax$		&	1.78	& 1.91	& 	1.96		& 1.94 	& 1.95	\\
$\ithreemax$ 		&	2.62	& 2.76	& 	2.81		& 2.82 	& 2.77 \\
\hline\hline
\end{tabular}
\caption{\label{tab:results}
`Truth-level' numerical calculations of the generalised CHSH and the CGLMP expectation values in $H\rightarrow WW^*$ decays,
as a function of the minimum invariant mass of either of the $W$ bosons. 
}
\end{table}

The numerical results in Table~\ref{tab:results} 
show that as $m_W^<$ is increased, and the $W$ bosons approach their mass shell, 
the value of $\itwomax$ approaches two.
However, that classical limit is not exceeded 
so no experimental Bell-inequality violation would be expected if using the 
generalised CHSH operator~\eqref{eq:chshoperatorspin}.
These findings are consistent with analytical and numerical results for this operator performed previously for 
entangled states of non-relativistic spin-one systems and of relativistic spin-one bosons in the narrow width approximation~\cite{Caban_2008}. 

The results for the CGLMP inequality are also shown in Table~\ref{tab:results}.
In this case the expectation values for all values of $m_W^<$ are well in excess of the classical limit of 2, and as large as $2.82$.
This is close to the largest possible value in non-relativistic quantum mechanics for a maximally entangled 
state\myendnote{See Ref.~\cite{Collins_2002}.  A slightly larger extremal value of $1+\sqrt{11/3}\approx2.9149$
can be achieved for other states that are not maximally entangled~\cite{Acin_2002}.} which is $4/(6\sqrt{3}-9)\approx2.8729$.
This result confirms that near-maximal violation of the CGLMP inequality is achieved despite relativistic and finite-width corrections.

In any real experiment corrections will be needed to account for detector acceptance and efficiency effects, for backgrounds, 
and for indeterminacies in the reconstructed $W$ bosons' rest frames.
The appropriate selections and corrections will vary from experiment to experiment,
and so only some initial estimates of their approximate magnitudes are considered here.

At a $pp$ collider such as the LHC one might employ kinematic methods similar to those of the  
`$M_{\rm T2}$-assisted on-shell' method~\cite{Cho_2009} to estimate the unobserved neutrino four-momenta. 
Here we provide a first estimate the effect of such reconstruction effects by independently\myendnote{
We neglect the kinematic constraints from the known masses and from the measured missing transverse momentum 
which could be used to constrain the overall set of momenta.}
smearing each of the three spatial components of 
the momentum of each of the two $W$ bosons with a Gaussian with two 
different values of the width parameter chosen to illustrate the size of the effect on the measurement.

\begin{table}
\setlength{\tabcolsep}{6pt}
\begin{tabular}{ l c  c  c c} 
\hline\hline
Expt. Assumptions			& Truth  	&	`A'		& `B'			& `C'		\\
\hline\hline
Min $p_T(\ell)$	[GeV]		& 0		& 	5 		&  	20	 	& 20		\\
Max $|\eta (\ell)| $			& ---		& 	2.5 		& 	2.5		& 2.5		\\
$\sigma_\mathrm{smear}$ [GeV]	& 0	& 	5		& 	5		& 10		\\
\hline\hline
Number of events			&  34.3k	& 	19.7k 	&  	6.5k	& 5.4k \\
Fraction of events			& 0.48 	&	0.27		&  	0.090	& 0.075	\\
$\ithreemax$				& 2.62	& 	2.40	 	& 	2.75	& 2.16	\\
Signif. $(\ithreemax-2)$		& $11.7\sigma$		& $5.2\sigma$ & $5.3\sigma$	& $1.0\sigma$ \\
\hline\hline
\end{tabular}
\caption{\label{tab:cuts}
Sensitivity of the CGLMP expectation value to experimental selection and resolution
for three different sets of experimental assumptions: `Truth', `A' and `B'. 
Rows 2--4 show the experimental cuts applied respectively 
to the lepton transverse momentum, and their pseudorapidity, 
and the smearing parameter for the reconstructed $W^{(*)}$ boson rest frames. 
Rows 5--8 show for an integrated $pp$ luminosity of $\intlumi$: the number and the fraction of the $H\rightarrow \ell^+\nu\ell^-\bar\nu$ events 
passing the selection, including the same-family lepton veto; 
the value of the measured CGLMP expectation value; and finally, for the same integrated luminosity, the statistical significance 
by which the CGLMP expectation value exceeds the classical limit of 2.
}
\end{table}

In Table~\ref{tab:cuts} we show results for three different experimental scenarios. 
The first `Truth' scenario shows the idealised results without modelling any experimental 
acceptance requirements or resolution effects.
Scenario `A' includes a resolution on the $W^{(*)}$ bosons' reconstructed momenta, 
modelled by a Gaussian of width 5\,GeV in each Cartesian direction, 
and also lepton acceptance requirements similar to those required 
employed by the LHC general-purpose detectors.
Scenario `B' has tighter lower bound of 20\,GeV on the leptons' transverse momenta, 
similar to that required by the leptonic triggers of the LHC detectors.
This reduces the number of events, but also has the effect of increasing the 
expectation value $\ithreemax$ as calculated from those remaining events. 
In a real experiment any bias introduced by the selection 
could be corrected by applying a similar selection requirement to
the angular integrals (\eqref{eq:Ax}-\eqref{eq:Az} and \eqref{eq:A4}-\eqref{eq:A5})
from which the density matrix coefficients are extracted,
but that correction is not applied here so that the magnitude of the uncorrected effect can be seen.
Scenario `C' is like `B' but with an increased smearing of the $W$ boson momentum to 
10\,GeV in each component. 
In each case all values of the $W^{(*)}$ bosons' masses are permitted, 
and the experimental backgrounds are neglected.  

The statistical significances\myendnote{The significance is here defined to be $(\ithreemax-2)/\sigma_\mathcal{I}$, 
where $\sigma_\mathcal{I}$ is the standard error of the mean of $\ithreemax$.}
by which the simulated values of $\ithreemax$ exceed the classical bound 
were calculated for  $pp$ integrated luminosity of $\intlumi$, and are also shown in Table~\ref{tab:cuts}. 
A significance of about $12 \sigma$ is found in the idealised `Truth' simulation,
falling to about $5\sigma$ for experimental scenarios `A' and `B' and to approximately 1$\sigma$ for scenario `C'.
The large significance for the idealised simulation demonstrates that the LHC has already produced sufficient
numbers of Higgs bosons to perform the measurements if experimental considerations could be neglected. 
The wide range of significances for different scenarios shows that making this measurement at the LHC
is likely to be experimentally demanding, and to require a careful study of -- and optimisation of -- these sorts of experimental effects.


\section{Discussion}

The near-maximal violation of the CGLMP inequality in numerical simulations 
motivates the more detailed study of Bell violation in Higgs boson 
decays at the LHC and also at future colliders.
We note that unlike in the case of e.g. $t\bar{t}$ production, $H\rightarrow WW^*$
involves an intermediate state comprising only a single narrow-width scalar -- the Higgs boson. 
This means that the spin density matrix $\rhoww$ of the $W$ boson pair 
does not depend on how the Higgs bosons have been produced. 
The expectation values of the Bell observables and the overall method are therefore the same for any Higgs boson 
whether produced in e.g. gluon-gluon fusion, in Higgsstrahlung, or vector-boson fusion. 

The overall method will translate directly to Higgs bosons produced at other high-energy accelerators, 
regardless of the types of particles $(p, e, \mu, \ldots)$ that are collided. 
At a future $e^+e^-$ collider of particular interest would be the $Z+H$ Higgsstrahlung production process, 
for which the Higgs signal events could be selected using the invariant mass of the object(s) recoiling against the $Z$ boson.
Such a selection would cleanly remove backgrounds such as those from $t\bar{t}$ and non-resonant $WW^*$ 
that would need to be accounted for at a $pp$ collider such as the LHC.  

An experimental observation of CGLMP violation in $H\rightarrow WW^*$ decays
would provide a striking conflict with realism deep within the regime in which field theory is expected to reign. 
Furthermore, since the $W^{(*)}$ bosons from the $H\rightarrow WW^*$ decay separate at relativistic speeds
they have a mixture of space-like and time-like separations at decay, 
so observation of Bell violation in this system might have implications for 
tests of causality and local realism -- again at these extreme length, time and energy scales.
If, by contrast, experiments are found to be unable to measure a Bell inequality violation where one is expected, 
then this would be an even more surprising and consequential result. 


\section{Conclusion}\label{sec:conclusion}
We have outlined methods by which two Bell inequalities --- a generalised CHSH inequality and the CGLMP inequality 
--- may be tested experimentally in $H\rightarrow WW^*$ decays, using the 
spin-analysing nature of the weak decays. 
Numerical simulations, agreeing with previous analytical and numerical work for pairs of spin-one bosons, 
suggest that one cannot expect to observe violation of the generalised CHSH inequality in this process.
By contrast the CGLMP inequality, the tightest inequality for pairs of three-state systems, 
is expected to be near-maximally violated in $H\rightarrow WW^*$ decays. 

The method described offers the opportunity to test Bell inequalities in the quantum field theory regime, 
providing prospects for experimental tests far removed from the length-scales, time-scales and energies of existing measurements.  
The many orders of magnitude difference in such scales provides ample scope for unexpected experimental results.

Similar test can be performed across a range of production channels (gg fusion, VBF, Higgsstrahlung, \ldots) 
and at any type of collider ($pp$, $ee$, $ep$, $\mu\mu$, \ldots) that produces Higgs bosons, 
since the narrow scalar $H$ does not retain information about its production mechanism.
The experimental challenges in each collider and in each production mechanism will differ, motivating dedicated studies of each. 

Numerical simulations suggest that, provided that experimental resolutions and selection effects can be controlled, 
then statistically significant violations of the CGLMP inequality might be observable by the LHC 
experiments using datasets comparable to those already collected.
\par
\par
~
\par
The author is grateful to Mateus Ara\'ujo, Claire Gwenlan, Chris Hays, Simon Saunders and to two anonymous referees for their comments on earlier versions of this manuscript.  
The author is funded through STFC grants ST/R002444/1 and ST/S000933/1. 


\bibliographystyle{JHEP}
\bibliography{HwwBell}

\end{document}